\documentclass[conference]{IEEEtran}
\IEEEoverridecommandlockouts

\usepackage{cite}
\usepackage{url}
\usepackage{amsmath,amssymb,amsfonts} 
\usepackage{algorithmic}
\usepackage{graphicx}
\usepackage{textcomp}
\usepackage{multirow} 
\usepackage{booktabs} 
\usepackage[hidelinks]{hyperref} 

\def\BibTeX{{\rm B\kern-.05em{\sc i\kern-.025em b}\kern-.08em
    T\kern-.1667em\lower.7ex\hbox{E}\kern-.125emX}}

\usepackage{fancyhdr} 


\begin{document}

\title{Classification of ADHD and Healthy Children Using Multi-Band and Spatial Features of EEG}

\author{
\IEEEauthorblockN{Md Bayazid Hossain}
\IEEEauthorblockA{\textit{Computer Science and Engineering} \\
\textit{Pabna University of Science and} \\ \textit{Technology,}
Pabna 6600, Bangladesh.\\ Email: bayazid204@gmail.com}
\and
\IEEEauthorblockN{Md Anwarul Islam Himel}
\IEEEauthorblockA{\textit{Computer Science and Engineering} \\
\textit{Pabna University of Science and} \\ \textit{Technology,}
Pabna 6600, Bangladesh.\\ Email: himelhim31@gmail.com}
\and
\IEEEauthorblockN{Md Abdur Rahim}
\IEEEauthorblockA{\textit{Computer Science and Engineering} \\
\textit{Pabna University of Science and} \\ \textit{Technology,}
Pabna 6600, Bangladesh.\\Email: rahim@pust.ac.bd}
\and
\IEEEauthorblockN{Shabbir Mahmood}
\IEEEauthorblockA{\textit{Computer Science and Engineering} \\
\textit{Pabna University of Science and} \\ \textit{Technology,}
Pabna 6600, Bangladesh.\\ Email: shabbir.cse@pust.ac.bd}
\and
\IEEEauthorblockN{Abu Saleh Musa Miah}
\IEEEauthorblockA{\textit{Computer Science and Engineering} \\
\textit{The University of Aizu,} \\
Aizuwakamatsu, Japan.\\
Email: abusalehcse.ru@gmail.com}
\and
\IEEEauthorblockN{Jungpil Shin}
\IEEEauthorblockA{\textit{Computer Science and Engineering} \\
\textit{The University of Aizu,} \\
Aizuwakamatsu, Japan.\\
Email: jpshin@u-aizu.ac.jp}
}

\maketitle

\begin{abstract}
Attention Deficit Hyperactivity Disorder (ADHD) is a common neurodevelopmental disorder in children, characterized by difficulties in attention, hyperactivity, and impulsivity. Early and accurate diagnosis of ADHD is critical for effective intervention and management. Electroencephalogram (EEG) signals have emerged as a non-invasive and efficient tool for ADHD detection due to their high temporal resolution and ability to capture neural dynamics. In this study, we propose a method for classifying ADHD and healthy children using EEG data from the benchmark dataset. There were 61 children with ADHD and 60 healthy children, both boys and girls, aged 7 to 12. The EEG signals, recorded from 19 channels, were processed to extract Power Spectral Density (PSD) and Spectral Entropy (SE) features across five frequency bands, resulting in a comprehensive 190-dimensional feature set. To evaluate the classification performance, a Support Vector Machine (SVM) with the RBF kernel demonstrated the best performance with a mean cross-validation accuracy of 99.2\% and a standard deviation of 0.0079, indicating high robustness and precision. These results highlight the potential of spatial features in conjunction with machine learning for accurately classifying ADHD using EEG data. This work contributes to developing non-invasive, data-driven tools for early diagnosis and assessment of ADHD in children.
\end{abstract}


\section{Introduction}
ADHD is a neurodevelopmental disorder marked by persistent patterns of inattention, hyperactivity, or impulsivity that interfere with daily functioning. These behaviours frequently occur across various settings, such as home, school, and work, and go beyond typical behaviour for most individuals\cite{b1}. Children with ADHD often struggle with sitting
still, planning ahead, completing tasks, or maintaining full awareness of their surroundings. The prevalence of ADHD is estimated to be around 12.1\% in boys and 3.9\% in girls\cite{b2}. ADHD is a prevalent condition frequently identified in children. Its symptoms typically start
in childhood and may persist into adolescence and adulthood. ADHD often occurs alongside other conditions, including conduct issues, learning disabilities, sleep disturbances, anxiety, or depression, which can complicate its diagnosis and treatment\cite{b3}.
For individuals with ADHD, symptoms often impact daily functioning. These symptoms can create challenges in completing tasks, disrupt school, work, and other activities, and put a strain on social relationships. Children with ADHD face a higher likelihood of injuries, social difficulties,
family stress, and academic struggles. Adolescents and adults with ADHD are also at a greater risk of engaging in risky behaviors\cite{b4}.
EEG is a valuable technique that offers insights into the brain’s background activity and serves as an indicator of the foundation for cognition and behavior\cite{b5}. According to a previous study, EEG plays an important role in the evaluation of neural function in children with ADHD \cite{b6}. Therefore, it can be a useful gadget for investigating and diagnosing the abnormal behaviour of ADHD children. 
In this paper, the continuity of attention will be investigated for ADHD and control children and using this continuity a new approach will be introduced. In the following, firstly EEG recording will be described, then the methodology of analysis  will be introduced and finally the results will be demonstrated.

\section{Literature Review}

There are many studies which employed EEG analysis for diagnosing ADHD. Recent studies have extensively employed EEG signal analysis to investigate ADHD. In Armin Allahverdy et al. (2016) the utility of nonlinear EEG features, achieving 96.7\% accuracy in ADHD classification using visual attention tasks\cite{b7}.
Ali Ekhlasi et al. (2020) explore disrupted information flow in ADHD children compared to healthy peers using directed phase transfer entropy (dPTE) on EEG data. Significant differences in connectivity in delta, beta, and theta bands were also observed, particularly in frontal and occipital regions. These findings underline the disrupted effective connectivity in ADHD brains, providing insights into its neurological underpinnings\cite{b8}. Mohammad Reza Mohammadi et al. (2016) proposed a method using nonlinear features of EEG, such as fractal dimensions, approximate entropy, and Lyapunov exponent. Feature selection methods such as mRMR and DISR were employed to enhance classification accuracy. A Multi-Layer Perceptron (MLP) neural network achieved accuracies of 92.28\% and 93.65\% with mRMR and DISR methods, respectively\cite{b9}. Farhana et al. (2023) presented an emotion recognition method using EEG signals by extracting narrowband spatial features from theta, alpha, beta, and gamma bands. The method involves preprocessing EEG data, subband decomposition using Butterworth bandpass filters, and calculating short-term entropy and energy features. A common Spatial Pattern (CSP) was employed to derive discriminative features, which were classified using a Support Vector Machine (SVM) with a polynomial kernel. The proposed approach outperforms several existing methods in binary emotion classification (valence and arousal), achieving accuracies of 96.15\% for DEAP and 99.95\% for SEED datasets \cite{b10}.
In 2013, Nazhvani et al, used N2 and P2 peaks of ERP to diagnose ADHD and achieved 92.9\% accuracy \cite{b11}.
These findings collectively underscore the potential of advanced EEG analysis in developing objective and precise ADHD diagnostic tools.

\section{Dataset Description and Data Preprocessing}

\subsection{Dataset Description}
The study involved 61 children diagnosed with ADHD and 60 healthy controls, consisting of both boys and girls aged 7 to 12 years. ADHD diagnoses were confirmed by an experienced psychiatrist using DSM-IV criteria, and the children had been on Ritalin for up to six months. In contrast, none of the children in the control group had a history of psychiatric disorders, epilepsy, or high-risk behaviors\cite{b12}. EEG recordings were conducted following the 10-20 standard, utilizing 19 channels (Fz, Cz, Pz, C3, T3, C4, T4, Fp1, Fp2, F3, F4, F7, F8, P3, P4, T5, T6, O1, O2) at a sampling frequency of 128 Hz. The reference electrodes, A1 and A2, were positioned on the earlobes. Since visual attention is often impaired in children with ADHD, the EEG protocol was designed around a visual attention task. During this task, participants viewed images of cartoon characters and were asked to count the number of characters in each image. The number of characters varied randomly between 5 and 16, and the images were sufficiently large enough that they were clearly visible and easily countable. To have a continuous stimulus during the signal recording, each image was displayed immediately and uninterrupted after the child’s response. As a result, the total duration of the recording depended on the child’s response speed throughout the task.
\begin{figure}[h!]
    \centering
    \includegraphics[width=0.3875\textwidth]{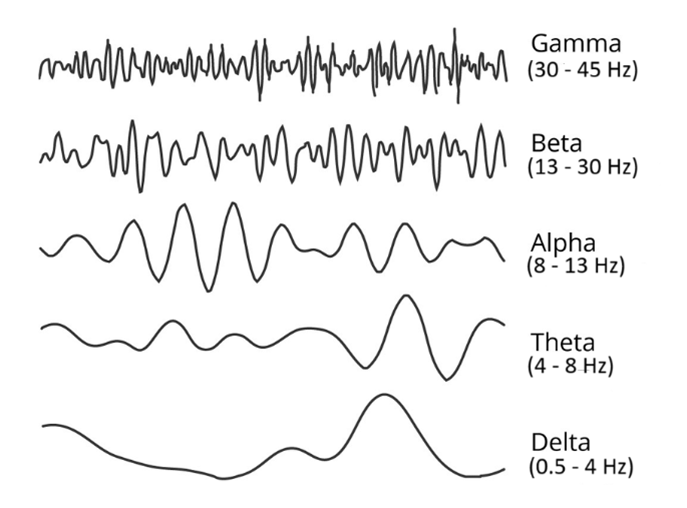} 
    \caption{The band description used in this study.}
    \label{fig:methodology-figure}
\end{figure}
\subsection{Data Preprocessing}
The EEG signals were digitized at the sampling rate of 128 Hz and recorded in the frequency range of 0.1-60 Hz. A bandpass Butterworth filter of  0.5 Hz to 50 Hz with order five was applied to continuous EEG data in order to eliminate artefacts. We have also tried using less than 50 Hz, which decreases the accuracy of the classification. Then, the signals are decomposed into multiple bands(Delta, Theta, Alpha, Beta, Gamma). We have used a bandpass Butterworth filter to decompose signal into sub-bands \cite{miah2017motor,miah2019motor,miah2019eeg,kibria2020creation_miah,miah2020motor}. The sub-bands descriptions are given on \href{fig 1}{fig 1}. For each sub-band, an average number of 29 epochs of 1280 points (10 seconds) with 640 points 50\% overlap were selected for each subject sub-band. Trial lengths exceeding 10 seconds are usually employed to analyze slower dynamics in the signal or to achieve higher frequency resolution\cite{b15}. The resulting data shape being \(\textstyle{(61 + 60)} \) subjects \(\textstyle{\times} \) 5 sub-bands \(\textstyle{\times} \) average 29 epochs \(\textstyle{\times} \) 19 channels \(\textstyle{\times} \) 1280 points.

\section{Methodology}
The proposed methodology for ADHD classification using EEG signals involves several key stages given in \href{fig 2}{fig 2}, starting from preprocessing to final classification. Initially, preprocessed EEG trials are subjected to subband decomposition, where the EEG signals are divided into multiple frequency bands (delta, theta, alpha, beta, gamma) to isolate distinct rhythmic components of brain activity. For each subband, two essential features, Power Spectral Density (PSD) and Spectral Entropy, are extracted. PSD represents the distribution of power across various frequency components, while spectral entropy quantifies the randomness or irregularity of the signal, both of which are crucial in distinguishing ADHD characteristics. 
\begin{figure}[h!]
    \centering
    \includegraphics[width=0.3875\textwidth]{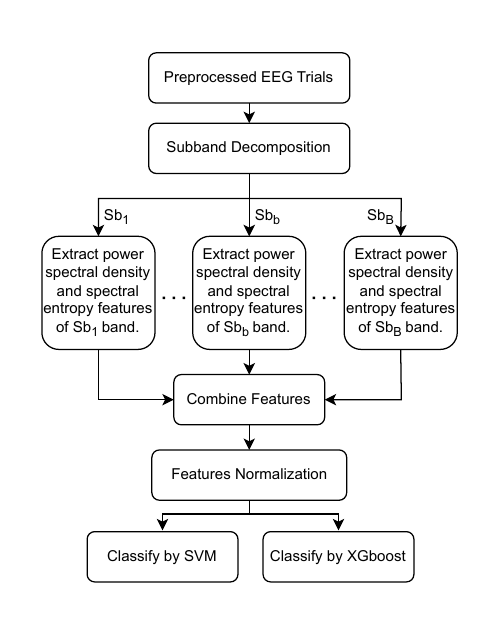} 
    \caption{The methodology figure.}
    \label{fig:methodology-figure}
\end{figure}
Once the features are extracted from all subbands, they are combined into a single feature set to enhance the robustness of the model by incorporating information from multiple subbands. The combined feature set is then passed through a normalization step to scale the features and reduce the effect of varying magnitudes across different features. This step is critical for ensuring the fair contribution of all features during classification.

Finally, the normalized feature set is fed into two classifiers, Support Vector Machine (SVM) and XGBoost, to perform ADHD classification. SVM is a widely used classifier that finds an optimal hyperplane to separate different classes, while XGBoost is a scalable and efficient gradient-boosting framework known for its high performance in complex classification tasks. We applied different SVM kernels (linear, polynomial, sigmoid, and RBF) to determine the best fit for the EEG data. The RBF kernel outperformed others, achieving 98.2\% accuracy, effectively capturing nonlinear patterns in brain activity. Using both classifiers, the methodology ensures that the best possible classification performance can be achieved for the ADHD data set.

\subsection{Feature Extraction}
In the previous study, they used non-linear features to classify ADHD. Here, we have used two spatial features: band power and spectral entropy. PSD helps analyze brain activity by measuring the power distribution across different frequency bands, which is known to differ in ADHD children\cite{b13}. And Spectral Entropy quantifies signal randomness, reflecting cognitive control and brain organization. Since ADHD is linked to altered neural oscillations and irregular brain activity, these features provide valuable insights for classification\cite{b14}. First of all, we found power spectral density(PSD) using Eqn\eqref{eq:psd}. From the PSD, the calculate band power Eqn\eqref{eq:bandPower}. And using Eqn\eqref{eq:bandEnergy}, calculate the band energy feature. After extracting features of all bands, then normalize the data using Eqn\eqref{eq:normalize}.

\begin{equation}
\label{eq:psd}
PSD(f)=\frac{1}{L}\sum\limits_{k=1}^{L}{\frac{1}{{{N}_{w}}\cdot {{F}_{s}}}\left| FF{{T}_{k}}(f) \right|_{{}}^{2}}
\end{equation}
Where:
\begin{itemize}
    \item  \(\textstyle {L} \): Number of segments the signal is divided into.
    \item \(\textstyle {{FFT}_{k}(f)} \): Discrete Fourier Transform (DFT) of the \(\textstyle {k-\text{th}}\)  segment at frequency \(\textstyle {f} \).
    \item  \(\textstyle {{N}_{w}}\): Number of samples in each segment.
    \item \(\textstyle {{F}_{s}}\): Sampling frequency of the signal.
    \item \( |\cdot|^2 \): Squared magnitude (to compute power).
    \item The factor \(\textstyle \frac{1}{{N}_{w}\cdot{F}_{s}}\): Normalizes the power to account for the sampling frequency and segment length.
\end{itemize}

\subsubsection{Band Power}
The band power calculates the total power of the signal within a specific frequency band \(\textstyle {[{f}_{low},{f}_{high}]}\). The equation is : 
\begin{equation}
    \label{eq:bandPower}
    {{P}_{band}}=\sum\limits_{f\in \left[ {{f}_{low}},{{f}_{high}} \right]}^{{}}{PSD(f)}
\end{equation}
Where:
\begin{itemize}
    \item  \(\textstyle PSD(f) \) : Power Spectral Density at frequency \(\textstyle {f}\),
    \item  \(\textstyle {[{f}_{low},{f}_{high}]}\) : Frequency range of the band (e.g., alpha, beta, etc).
\end{itemize}
\subsubsection{Band Entropy}
The band entropy is calculated using the normalized band-limited PSD. The formula for entropy is derived from Shannon's entropy:
\begin{equation}
    \label{eq:bandEnergy}
    {{E}_{band}}=\sum\limits_{f\in [{{f}_{low}},{{f}_{high}}]}^{{}}{p(f)\cdot \log (p(f))}
\end{equation}
Where:
\begin{itemize}
    \item  \(\textstyle p(f) = \frac{PSD(f)}{\sum\limits_{f\in [{{f}_{low}},{{f}_{high}}]}^{} {PSD(f)}}\) : Normalized PSD values within the band,
    \item  \(\textstyle {\sum}{p(f)} = 1\) : Ensures \(\textstyle {p(f)}\) is a probability distribution.
    \item A small value (e.g., \(\textstyle 1\times{10^{-8}}\) is added to \(\textstyle {p(f)}\) in the code to avoid \(\textstyle{log(0)}\).
\end{itemize}

\subsection{Feature Normalization}
The extracted features are normalized by the following formula: 
\begin{equation}
    \label{eq:normalize}
    z=\frac{x-\mu }{\sigma }
\end{equation}
Where:
\begin{itemize}

    \item  \(\textstyle {z} \): The normalized value.
    \item  \(\textstyle {x} \): The original feature value.
    \item  \(\textstyle {\mu} \): Mean of the feature values.
    \item  \(\textstyle {\sigma} \): The standard deviation of the feature values.
\end{itemize}
This feature normalization ensures that the normalized features have a mean of 0 and a standard deviation of 1.

\subsection{Classification}
\subsubsection{Classificaton By SVM}
The performs a comparative evaluation of different kernel functions (linear, rbf, poly, and sigmoid) for an SVM model. To evaluate, use 10-fold cross-validation. 
\begin{table}[htbp]
\caption{Comparison of Model Performance}
\centering
\begin{tabular}{|c|c|c|c|c|}
\hline
\textbf{Kernel} & \textbf{Mean Accuracy} & \textbf{Standard Deviation}\\
\hline
Linear & 94.90\% & 0.0216\\
\hline
\textbf{Rbf} & \textbf{99.20\%} & \textbf{0.0079} \\
\hline
Poly & 98.40\% & 0.0107\\
\hline
Sigmoid & 91.39\% & 0.0216\\
\hline
\end{tabular}
\label{tab:kernel_res}
\end{table}
For each kernel, we initialize an SVM model (SVC) with the specified kernel type, along with fixed regularization parameter \(C=1.0\), a scaling gamma value (gamma='scale').

The model's performance is assessed on the training data using cross-validation. For each kernel, here calculates the individual accuracy scores across the 10 folds, as well as the mean accuracy and standard deviation. The result of this experiment given on Table I. This analysis helps compare the effectiveness of different kernel functions in terms of their classification accuracy and variability.
\subsubsection{XGBoost Classifier}
The classifier with specific hyperparameters is tailored to train a robust model. The classifier is set to use 100 trees (n\_estimators) and a learning rate of 0.1, ensuring a balanced step size for weight updates during boosting. The maximum depth of each tree is limited to 6 (max\_depth), controlling the complexity of the model and preventing overfitting. Additionally, the training process uses 80\% of the samples (subsample) and 80\% of the features (colsample\_bytree) for each tree, introducing randomness to improve generalization. This configuration is designed to provide a balance between accuracy and generalization, making the model robust and effective.
\section{Experimental Results}
The experimental evaluation of the proposed methodology was conducted using two classifiers: SVM with an RBF kernel and XGBoost. The results indicate that both models achieved high performance, but SVM outperformed XGBoost in terms of accuracy and other metrics in Table~\ref{tab:accuracy_gr}.
\begin{table}[htbp]
\caption{Comparison of Model Performance}
\centering
\begin{tabular}{|c|c|c|c|c|}
\hline
\textbf{Methodology} & \textbf{Accuracy} & \textbf{Precision} & \textbf{Recall} & \textbf{F1-Score} \\
\hline
\textbf{SVM(rbf kernel)} & \textbf{99.20\%} & \textbf{0.98} & \textbf{0.99} & \textbf{0.99} \\
\hline
XGBClassifier & 96.92\% & 0.95 & 1.00 & 0.97 \\
\hline
\end{tabular}
\label{tab:accuracy_gr}
\end{table}

\begin{figure}[htbp]
\centering
\includegraphics[width=3.05in]{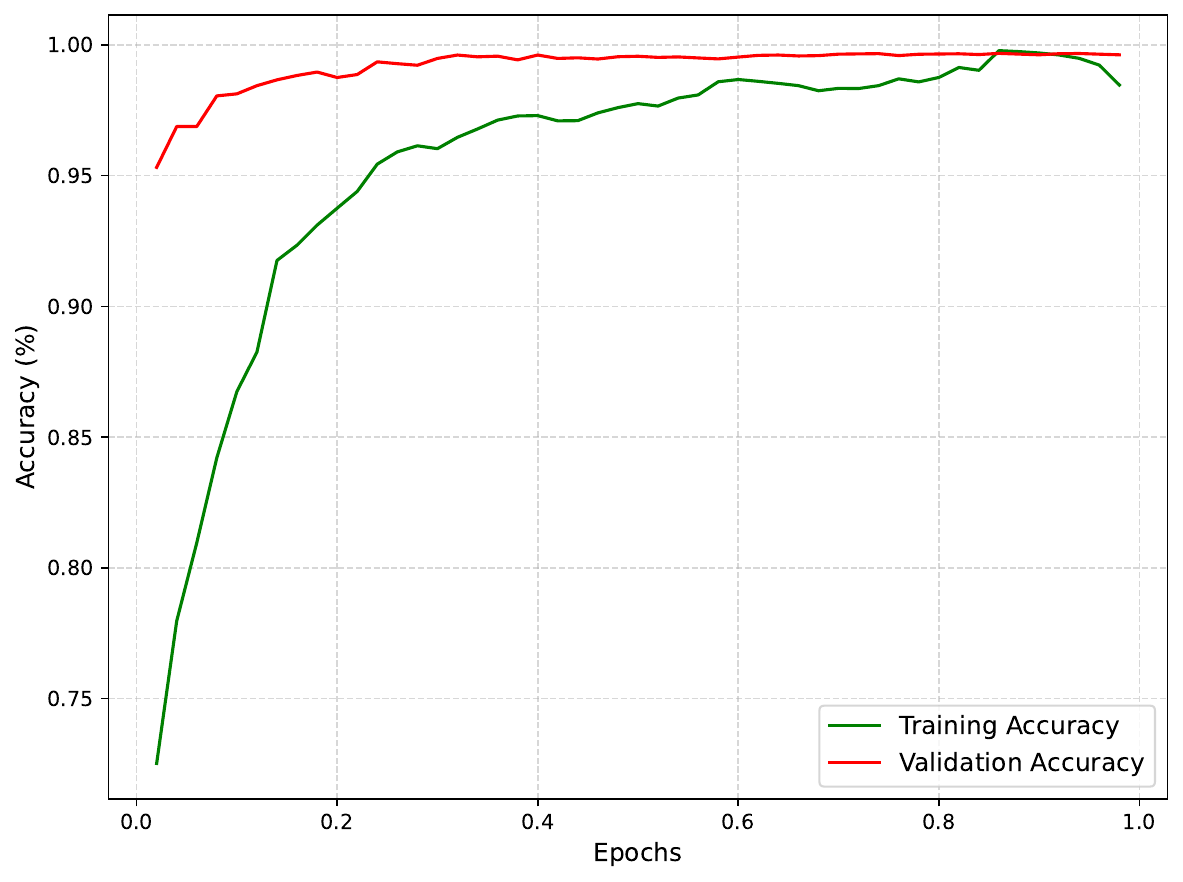}
\caption{Accuracy Curve}
\label{fig:accuracy_cr}
\end{figure}

Figure~\ref{fig:accuracy_cr} shows the accuracy curve for both training and validation phases across multiple epochs. It can be observed that the validation accuracy stabilizes after an initial rapid increase, reaching a maximum value close to 99\%, indicating that the model generalizes well to unseen data without significant overfitting.

SVM with an RBF kernel outperformed XGBoost, achieving higher accuracy (99.2\% vs. 96.92\%). This can be attributed to SVM’s ability to find an optimal decision boundary in high-dimensional feature spaces, which is crucial given the complexity of EEG signals. The RBF kernel effectively captures non-linear patterns in brain activity, making it well-suited for differentiating ADHD and healthy children. On the other hand, while XGBoost is powerful in handling structured data, it may not generalize as well to highly non-linear relationships in EEG data. Additionally, SVM's robustness against small dataset sizes and high-dimensional features likely contributed to its superior performance.\\
Overall, these results demonstrate the effectiveness of the proposed methodology, with SVM providing slightly superior performance in terms of accuracy and F1-score.

\section{Conclusions}
This study demonstrated that multi-band EEG features, combined with machine learning, achieve high accuracy (98.2\%) in classifying ADHD and healthy children. The use of Power Spectral Density and Spectral Entropy features proved to be highly effective in capturing ADHD-related neural patterns, reinforcing the potential of EEG as a non-invasive and objective diagnostic tool. The superior performance of SVM with an RBF kernel highlights its ability to handle the complex, non-linear nature of EEG data, making it a strong candidate for real-world applications.\\ Despite these promising results, some limitations should be considered. The size of the data set, though reliable for initial analysis, remains relatively small for a broader generalization. Future studies should validate the approach in larger and more diverse populations.

\vspace{12pt}

\end{document}